\documentclass[floatfix,lengthcheck,showpacs,amssymb,amsmath,amsfonts,twocolumn,nofootinbib,longbibliography]{revtex4-1}

\usepackage{lineno}
\usepackage{color}
\usepackage{graphicx}
\usepackage{float}
\usepackage{subfigure}
\usepackage{amsmath}
\usepackage{multirow}
\usepackage{tikz}
\usetikzlibrary{arrows,shapes,trees,decorations.pathreplacing}
\usepackage{import}
\usepackage{svn-multi}
\usepackage{hyperref}
\usepackage{gensymb}
\hypersetup{
    colorlinks=true,
    linkcolor=blue,
    filecolor=magenta,      
    urlcolor=cyan,
}
\DeclareUnicodeCharacter{2009}{\,} 
\usepackage[capitalise]{cleveref}
\crefname{figure}{Fig.\,}{Figs.\,}
\Crefname{figure}{Fig.\,}{Figs.\,}

\newcommand{\msun}{\ensuremath{\,M_{\odot}}}

\begin{document}

\title{Broad search for gravitational waves from subsolar-mass binaries through LIGO and Virgo's third observing run}

\author{Alexander H. Nitz}
\email{alex.nitz@aei.mpg.de}
\author{Yi-Fan Wang}
\affiliation{Max-Planck-Institut f{\"u}r Gravitationsphysik (Albert-Einstein-Institut), D-30167 Hannover, Germany}
\affiliation{Leibniz Universit{\"a}t Hannover, D-30167 Hannover, Germany}

\begin{abstract} 
We present a search for gravitational waves from the coalescence of binaries which contain at least one subsolar mass component using data from the LIGO and Virgo observatories through the completion of their third observing run.  The observation of a merger with a component below $1\,M_{\odot}$ would be a clear sign of either new physics or the existence of a primordial black hole population; these black holes could also contribute to the dark matter distribution. Our search targets binaries where the primary has mass $M_1$ between 0.1-100\,$M_{\odot}$ and the secondary has mass $M_2$ from 0.1-1\,$M_{\odot}$ for $M_1 < 20\,M_\odot$ and 0.01-1\,$M_{\odot}$ for $M_1 \ge 20\,M_\odot$. Sources with $M_1 < 7\,M_\odot, M_2 > 0.5\,M_\odot$ are also allowed to have orbital eccentricity up to $e_{10} \sim 0.3$. This search region covers from comparable to extreme mass ratio sources up to $10^4:1$. We find no statistically convincing candidates and so place new upper limits on the rate of mergers; our analysis sets the first limits for most subsolar sources with $7\,\msun< M_1 < 20\,M_{\odot}$ and tightens limits by $\sim 8\times$ $(1.6\times)$ where $M_1 > 20\,M_{\odot}$ ($M_1 < 7\,M_{\odot}$). Using these limits, we constrain the dark matter fraction to below 0.3 (0.7)$\%$ for 1 (0.5)\,$M_{\odot}$ black holes assuming a monochromatic mass function. Due to the high merger rate of primordial black holes beyond the individual source horizon distance, we also use the lack of an observed stochastic background as a complementary probe to limit the dark matter fraction. We find that although the limits are in general weaker than those from the direct search they become comparable at $0.1 \,M_{\odot}$.
\end{abstract}
\date{\today}
\maketitle

\section{Introduction}
Gravitational waves are now regularly observed by the ground-based observatories Advanced LIGO~\cite{TheLIGOScientific:2014jea} and Advanced Virgo~\cite{TheVirgo:2014hva}. Their accomplishments include more than 90 observed binary black hole (BBH) mergers~\cite{LIGOScientific:2021djp,Nitz:2021zwj,Olsen:2022pin} and a handful of binary neutron star~\cite{TheLIGOScientific:2017qsa,Abbott:2020uma} and neutron star -- black hole mergers~\cite{LIGOScientific:2021qlt}. These observations provide a wealth of knowledge for understanding the population of stellar-mass black holes~\cite{LIGOScientific:2021psn}, which may have arisen through standard stellar evolution~\cite{Mandel:2021smh}. 
Field evolution \cite{Kalogera:2006uj,Dominik:2014yma,Belczynski:2016obo,Giacobbo:2018etu} and dynamical formation channels \cite{Antonini:2015zsa,Rodriguez:2015oxa,Rodriguez:2016kxx,Park:2017zgj,Fragione:2018vty,Gerosa:2021mno} for compact binaries have been proposed to describe this process. Several observations though have challenged current understanding of stellar formation; these include GW190521 \cite{LIGOScientific:2020ufj,LIGOScientific:2020iuh}, the observation of a merger that includes a black hole that may be in the ``upper mass gap" caused by pair-instability supernovae~\cite{Woosley:2016hmi,Marchant:2018kun,Stevenson:2019rcw}.  
Observations also confirm the existence of compact objects with secondary mass $1-3\msun$ (e.g. GW190814) \cite{Abbott:2020khf,LIGOScientific:2021psn}. 
It has been proposed that such events may be composed of primordial black holes (PBHs)~\cite{DeLuca:2020sae, Clesse:2020ghq,Vattis:2020iuz,Jedamzik:2020omx,Wang:2021iwp}. Studies have also shown that the current population of observed binary black holes is compatible with and may include contributions from a population of merging PBHs~\cite{Franciolini:2021tla,DeLuca:2021wjr,Chen:2021nxo}. 

PBHs are hypothesized to form by direct collapse of overdensity in the very early Universe \cite{1966AZh....43..758Z,10.1093/mnras/152.1.75}, and have implications for several astrophysical and cosmological scenarios \cite{Clesse:2017bsw}; these include seeding the first galaxies and the formation of supermassive black holes~\cite{Duechting:2004dk,Khlopov:2008qy}, explaining the recent excess power detected by pulsar timing arrays \cite{DeLuca:2020agl,Kohri:2020qqd,Vaskonen:2020lbd}, and most intriguingly, as a candidate for dark matter~\cite{10.1093/mnras/152.1.75}. A variety of astrophysical observations have put constraints on PBH abundance (for a review see e.g.~\cite{Carr:2020gox}).
So far, dark matter has evaded all direct searches based on experiments on the earth, including the hunt for weakly interacting massive particles (WIMPs) \cite{PandaX-4T:2021bab} and axion dark matter \cite{ADMX:2021nhd}. The interest in macroscopic dark matter candidates such as PBHs has been revived in light of the discovery of black hole mergers by LIGO and Virgo~\cite{Bird:2016dcv,Sasaki:2016jop,Clesse:2016vqa}.

Among the gravitational-wave catalog, GW190814 is a source of particular interest due to its mass ratio (q$\sim$10) and low spin~\cite{Abbott:2020khf}. Its spin is consistent with a merger of primordial origin~\cite{Carr:2019kxo,Clesse:2020ghq}, however more mundane explanations are also possible~\cite{Liu:2020gif}. 
Like GW190814, GW190425 \cite{LIGOScientific:2020ibl}, GW191219, GW200105, GW20015 \cite{LIGOScientific:2021qlt}, and GW200210 all contain secondary components with mass less than $3\msun$~\cite{LIGOScientific:2021psn}, which is lighter than the lower limit observed from X-ray binaries~\cite{Bailyn:1997xt,2010ApJ...725.1918O,2011ApJ...741..103F}.
If there exists a distribution of binaries composed of PBHs, the population could be convincingly demonstrated by the observation of a similar source with a subsolar-mass black hole secondary. The detection of a subsolar-mass black hole would confidently establish the existence of PBHs~\cite{Sasaki:2018dmp,Carr:2020gox,Green:2020jor} or other exotic physics able to produce subsolar-mass black holes~\cite{Shandera:2018xkn,Kouvaris:2018wnh,Dasgupta:2020mqg} due to their inability to form through standard stellar evolution~\cite{minmass1,minmass2}.

\begin{figure}[tb!]
    \centering
    \includegraphics[width=\columnwidth]{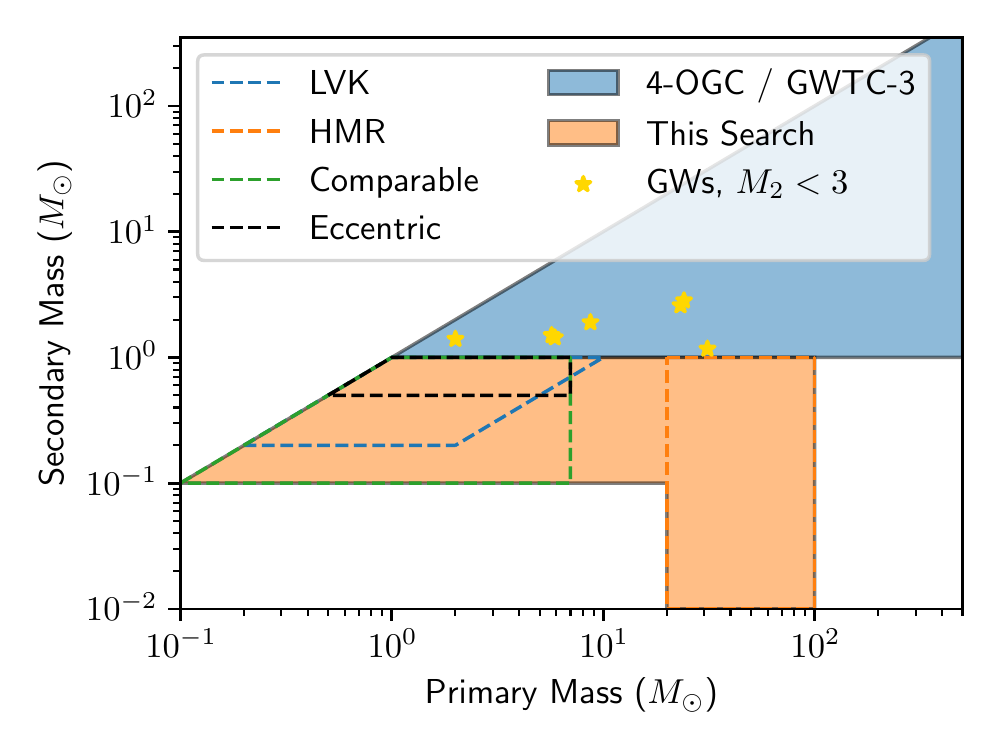}
    \caption{The primary and secondary (redshifted) masses of the sources searched by our analysis (orange), 4-OGC/GWTC-3 (blue)~\cite{Nitz:2019hdf,LIGOScientific:2018mvr}, and the subsolar mass LVK search (blue dotted)~\cite{LIGOScientific:2021job}. The boundaries where our search also includes sources with eccentricity up to 0.3 (black dashed), along with the boundaries of our prior comparable mass (green dashed) \cite{Nitz:2021vqh,Nitz:2021mzz} and high-mass ratio (orange dashed) searches \cite{Nitz:2020bdb} are shown. For comparison, we show the masses of reported gravitational-wave sources closest to our search
    region (yellow stars); if any of these is primordial in origin, the population may extend into our search region. 
    }
    \label{fig:searches}
\end{figure}

In this letter, we conduct a search to answer if LIGO and Virgo have observed any mergers that include a subsolar-mass secondary using the most recent data from the three completed observation runs (O1-O3)~\cite{Vallisneri:2014vxa,Abbott:2019ebz}. Previous work has directly searched for comparable mass subsolar mass sources up to the first half of the third observation run (O3a)~\cite{Nitz:2021mzz,Nitz:2021vqh,LIGOScientific:2021job,LIGOScientific:2019kan,LIGOScientific:2018glc,Phukon:2021cus}, and high-mass-ratio sources through the second observation run (O2) \cite{Nitz:2020bdb}. We perform a broad analysis designed to be sensitive to sources where the primary mass $M_1$ is 0.1-100~$M_\odot$ while the secondary can range from $0.01 - 1\msun$ for $M_1 > 20\msun$ and $0.1-1\msun$ for $M_1<20\msun$. Fig.~\ref{fig:searches} shows the boundaries of this search and how it compares to previous analyses; our search encompasses the mass ranges of prior searches and for the first time includes the full range where the primary mass is $7-20 M_\odot$. The most significant candidate in our search had a false alarm rate of $\sim$ 1 per 1.8 years. Due to the time searched, we consider this a null observation. With the inclusion of the most recent data, our limits on the rate of mergers are  $\sim8\times$ more stringent than our prior high-mass-ratio search~\cite{Nitz:2020bdb} and $1.6\times$ the comparable mass search~\cite{Nitz:2021vqh,Nitz:2021mzz} which only included up to O2 and O3a, respectively. We find the merger rate of 0.5-0.5 (1-1)~\msun~ binaries is $< 4400 (700)~\mathrm{Gpc}^{-3} \mathrm{yr}^{-1}$ and for 1 - 20~\msun~ sources is $< 65~\mathrm{Gpc}^{-3} \mathrm{yr}^{-1}$. 

Our non-detection constrains astrophysical population models which predict the binary PBH merger rate. For a fiducial monochromatic mass population, we find that the dark matter mass fraction of PBHs is $\le6\%(0.3\%)$ for component mass $0.1(1)\msun$. For a two-point mass function where the primary mass is fixed to be $37\msun$ (approximately the average value for all detections), the mass fraction is $\le3\%(0.03\%)$ for secondary mass $0.01(1)\msun$.
Because the stochastic gravitational-wave background ~\cite{Allen:1996vm,Allen:1997ad,Romano:2016dpx} provides information about the population of unresolved, high-redshift sources and the PBH merger rate increases with redshift beyond the detection horizon for individual sources, we also use the non-detection of a stochastic background in O3 data~\cite{KAGRA:2021kbb} to infer additional constraints.

\begin{figure*}[tb!]
    \centering
    \includegraphics[width=2\columnwidth]{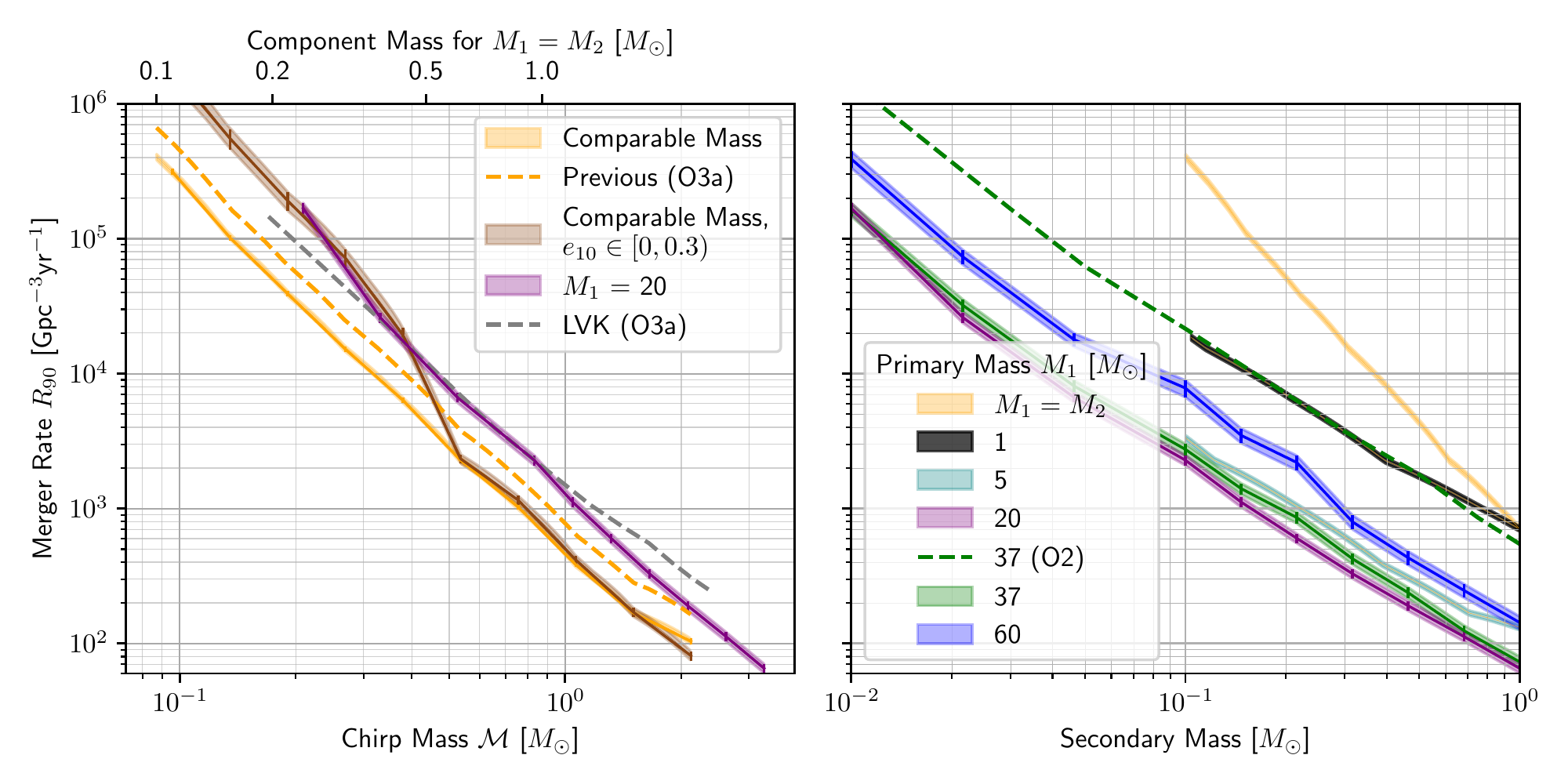}
    \caption{The $90\%$ upper limit on the rate of mergers from the null detection in our direct search for compact-binary mergers. Left: The upper limit as a function of chirp mass. For sources with comparable mass components, the search sensitivity can be approximated as only a function of the chirp mass.
    The most recent results from our prior analysis (orange dashed)~\cite{Nitz:2021vqh} and the LVK (gray dashed)~\cite{LIGOScientific:2021job} are shown for reference. Our limits are up to 3x more constraining than the most recent LVK results, and have improved by $\sim60\%$ over our prior limits. For comparison, a 20:1 mass ratio system is also shown; high-mass-ratio sources will have reduced sensitivity in comparison to comparable mass at the same chirp mass due to the effect of higher-order gravitational-wave modes and the merger moving into the sensitive frequency band of the observatories. Right: The upper limit as function of the secondary mass for a selection of primary masses. Our new limits for $M_1=37\msun$ (green solid) have improved by 8x over the prior state-of-the art (green dashed)~\cite{Nitz:2020bdb}.  The one sigma Monte-Carlo statistical uncertainty is shown with shading.}
    \label{fig:rate}
\end{figure*}

\section{Subsolar Mass Search Results}

Our search is conducted using the compact-binary analysis included in the open-source PyCBC toolkit~\cite{pycbc-github}. The analysis identifies candidates~\cite{Usman:2015kfa,Allen:2005fk,Davies:2020}, checks for consistency between the data and astrophysical sources~\cite{Allen:2004gu, Nitz:2017svb, Nitz:2017lco}, and assesses each candidate's statistical significance. Our analysis targets the parameters space shown in Fig.~\ref{fig:searches} and is configured similarly to our previously conducted analyses for comparable mass sources in Refs.~\cite{Nitz:2021vqh,Nitz:2021mzz} and high-mass-ratio binaries in Ref.~\cite{Nitz:2020bdb}. 

To detect sources, our matched-filter based analysis requires a model of the expected gravitational-wave signal. We model the signal using the TaylorF2 waveform template derived from the post-Newtonian expansion for noneccentric sources~\cite{Sathyaprakash:1991mt,Droz:1999qx,Blanchet:2002av,Faye:2012we} where $M_1 < 7~\msun$ and EOBNRv2~\cite{Pan:2011gk}, which includes a model of the merger and ringdown phase of the signal for higher masses, elsewhere. We use TaylorF2e~\cite{Moore:2019a,Moore:2019vjj,Moore:2018} to model eccentric sources up to $e_{10} \sim 0.3$ where $0.5 \msun < M_1 < 7\msun, M_2 > 0.5\msun$ and $e_{10}$ is the binary eccentricity at a reference gravitational-wave frequency of 10 Hz. Our template models only include the dominant mode of the gravitational-wave signal. A discrete set of template waveforms is selected using a stochastic placement algorithm~\cite{Harry:2009ea} to ensure the signal-to-noise ratio (SNR) loss due to template bank density is $3-5\%$ on average. To control for computational cost, the search analyzes data from a minimum of 20 Hz, but this frequency is raised independently for each template to ensure its duration is no more than 512s where $M_1 < 7~\msun$ and 60s everywhere else. 

The public LIGO and Virgo dataset now contains data from all three observing runs through 2021 which amounts to 1.2 years of multi-detector time~\cite{Vallisneri:2014vxa,Abbott:2019ebz}. Our analysis finds no convincing gravitational-wave detections; the most significant candidate has a false alarm rate of 1 per 1.8 years which is consistent with a null observation given the observation time. Using a null detection, we can set an upper limit at $90\%$ confidence on the rate of mergers (shown in Fig.~\ref{fig:rate}) as a function of the binary parameters using 

\begin{equation}
    R_{90}(m_{1,2}) = \frac{2.3}{VT(m_{1,2})},
    \label{eq:singlerate}
\end{equation}

where VT is the measured volume-time of the search~\cite{Biswas:2007ni}. We measure the VT of our analysis as a function of the component masses of a binary by empirically measuring the response of our analysis to simulated signal populations. Sources are assumed to have isotropic orientation and sky location in addition to a uniform distribution in volume. For comparable mass sources we use TaylorF2(e) as a source model. For $M_1 > 7$ where mass ratio can be up to $10^4$, we use the EOBNRv2HM model which includes sub-dominant modes of the gravitational-wave signals~\cite{Pan:2011gk,lalsuite}; this allows us to account for the effect of neglecting sub-dominant modes in our search. In addition, we assume that PBHs will have negligible spin which is consistent with the predicted spin distribution~\cite{Chiba:2017rvs,DeLuca:2019buf,DeLuca:2020bjf,Mirbabayi:2019uph,Postnov:2019pkd}. For large total mass binaries, $M_1 + M_2 > \sim 35-80 \msun$, however, we note that non-negligible spin may be induced depending on the level of accretion~\cite{Franciolini:2022iaa}.

\begin{figure*}[tb!]
    \centering
    \includegraphics[width=2\columnwidth]{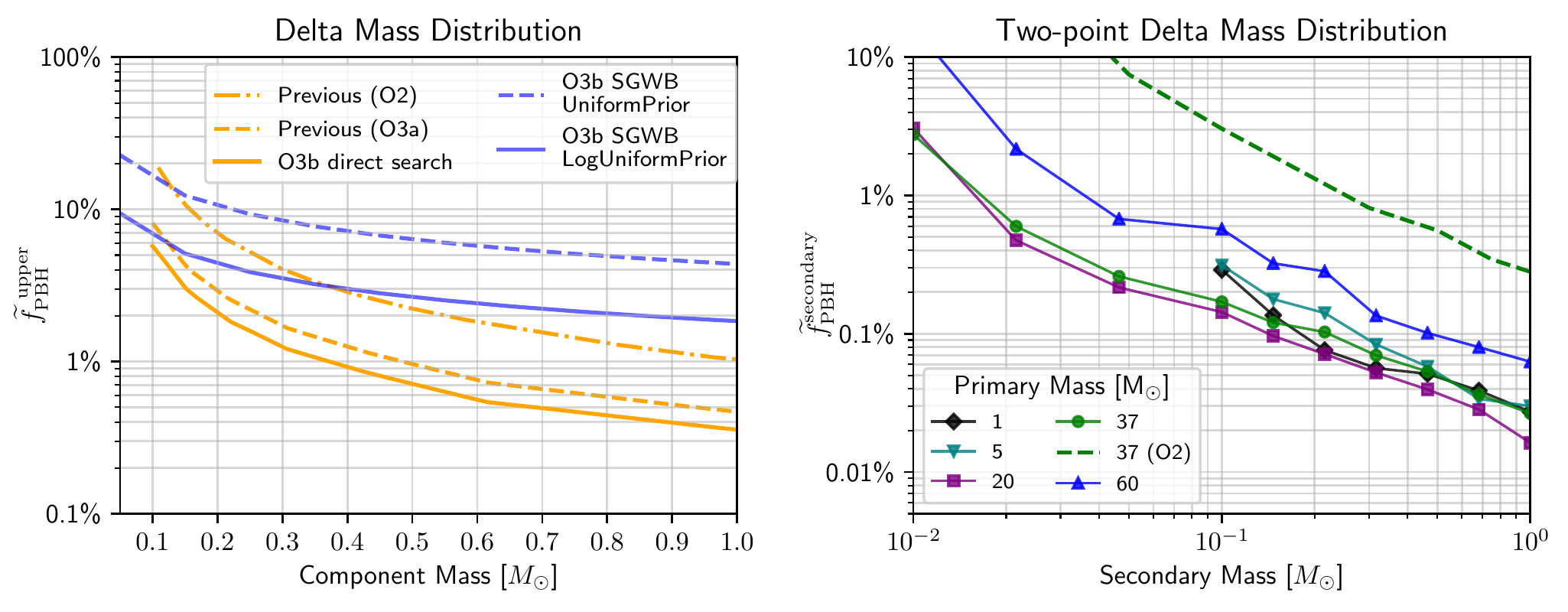}
    \caption{Left: $90\%$ upper limits on the fraction of dark matter accounted for by PBHs assuming a delta function mass distribution. The orange lines are constraints from our direct searches with data through O2, O3a, and O3b. The blue solid and dashed lines are constraints from the stochastic background.
    Right: $90\%$ upper limits on the abundance of PBHs for the secondary black hole assuming a two-point delta function for mass. To allow consistent comparison with previous work , we follow Ref.~\cite{Nitz:2021mzz} to fix the abundance of the primary mass to be $f^\mathrm{primary}_\mathrm{PBH} = 3\times10^{-3}$ which is obtained by fitting the observed black hole merger population\cite{Chen:2021nxo}. The solid lines are constraints for different representative primary mass. The dash line is the prior constraint based on data only through O2.}
    \label{fig:fpbh}
\end{figure*}

\section{Observational Constraints on Primordial Black Hole Dark Matter Contribution}

Limits on the observed merger rate can constrain the dark matter mass fraction of PBHs given a population model for the binary merger rate. 
We use the model in Refs.~\cite{Nakamura:1997sm,Sasaki:2016jop,Ali-Haimoud:2017rtz, Chen:2018czv}, which assumes a Poisson spatial distribution for PBHs when they initially form from large overdensity collapse in the early Universe. 
A nearby pair of black holes form a binary after decoupling from the cosmic expansion and then inspiral due to gravitational radiation.
Given a general mass distribution $P(m)$, the binary merger rate in units of $\mathrm{Gpc}^{-3} \mathrm{yr}^{-1}$ is
\begin{eqnarray}\label{eq:broadmassrate}
&R&(\widetilde{f}_\mathrm{PBH},m_{1/2},t) = 3\times10^6 \widetilde{f}_\mathrm{PBH}^2(0.7\widetilde{f}_\mathrm{PBH}^2 + \sigma_\mathrm{eq}^2)^{-\frac{21}{74}} \nonumber \\
&\times&(m_1m_2)^{\frac{3}{37}}(m_1+m_2)^{\frac{36}{37}}\mathrm{min}\left(\frac{P(m_1)}{m_1},\frac{P(m_2)}{m_2}\right) \nonumber \\
&\times&\left(\frac{P(m_1)}{m_1}+\frac{P(m_2)}{m_2}\right) \left(\frac{t}{t_0}\right)^{-\frac{34}{37}},
\end{eqnarray}
where $m_{1/2}$ are in unit of $\msun$, $t$ is the cosmic time and $t_0$ is the age of the Universe, both in units of years, $\sigma_\mathrm{eq}=0.005$ characterizes the variance of density inhomogeneity from dark matter at the mass density equality era. 
The effective mass fraction is defined to $\widetilde{f}_\mathrm{PBH}^{53/37} = S f_\mathrm{PBH}^{53/37}$ where $f_\mathrm{PBH}$ is the true mass fraction and the suppression term $S$ accounts for binary disruption after formation. Refs.~\cite{Raidal:2018bbj,Jedamzik:2020ypm,Hutsi:2020sol} showed the disruption by nearby PBH clusters can reduce the merger rate by orders of magnitude if PBHs account for nearly all dark matter but become negligible when $f_\mathrm{PBH}\lesssim\mathcal{O}(0.1\%)$.

In this letter we focus on a common fiducial mass distribution following \cite{Nitz:2020bdb, Nitz:2021mzz, Nitz:2021vqh} where the component masses are fixed to chosen values. Limits for arbitrary extended distributions can also be derived from our limits (see e.g. ~\cite{Kuhnel:2017pwq,Carr:2017jsz,Nitz:2021mzz}).

\subsection{Constraints from our direct search}

We first use our direct search limits to constrain the dark matter mass fraction from PBHs. Results are shown in \cref{fig:fpbh}. In the left panel the PBH mass function is assumed to be a delta distribution. For 1 and 0.1 $\msun$ reference masses, the fraction of dark matter composed of PBHs can not exceed $0.3\%$ and $6\%$, respectively. The right panel of \cref{fig:fpbh} shows constraints where we choose the mass function to be a two-point delta distribution. We fix $\tilde{f}_\mathrm{PBH}=0.3\%$ for the primary mass and constrain the $\tilde{f}_\mathrm{PBH}$ of the secondary PBH. This choice is motivated by Ref.~\cite{Chen:2021nxo} which fitted a PBH mass function and abundance with the current black hole merger observations~\cite{LIGOScientific:2021djp} assuming they are primordial in origin. We consider five representative primary component masses 1, 5, 20, 37 (approximately the average mass of all gravitational wave events) and 60 $\msun$, and secondary mass range in [0.01,1]~$\msun$. The previous constraint for $M_1 = 37\msun$, which included data only up to O2, are also shown for comparison. The O3b constraints improved over our previous by one order of magnitude due to upgrades of the advanced detectors for O3 and the longer duration of observation~\cite{LIGO:2021ppb,LIGOScientific:2021djp}

\subsection{Constraints from the stochastic background}

In addition to individually resolvable sources, the incoherent superposition of all binary PBH coalescences produces a stochastic background of gravitational waves \cite{Mandic:2016lcn,Wang:2016ana}. 
While the astrophysical binary black hole merger rate approximately follows the star formation rate peaking at redshift $\sim2$ \cite{Rosado:2011kv}, the PBH merger rate gets higher when redshift increases \cite{Nakamura:1997sm,Sasaki:2016jop}.
The search for an isotropic stochastic gravitational-wave background by the LVK did not find any significant excess energy in their O3 analysis~\cite{KAGRA:2021kbb}. 
We investigate the implication of this non-detection of a stochastic background for the PBH abundance.

The stochastic background is characterized by 
\begin{equation}
\Omega_\mathrm{GW} = \frac{\nu}{\rho_c}\frac{d\rho}{d\nu}   
\end{equation}
where $d\rho$ is the gravitational-wave energy density in a frequency bin $[\nu, \nu+d\nu]$, normalized by critical energy density for a flat Universe $\rho_c$.
Given the merger rate model for PBH binaries from Eq. \ref{eq:broadmassrate}, the stochastic background can be computed as
\begin{equation}\label{eq:omegagw}
\Omega_\mathrm{GW} = \frac{\nu}{\rho_c}\int \frac{R(z; \tilde{f}_\mathrm{PBH},m_{1/2})}{(1+z)H(z)}\frac{dE_\mathrm{s}}{d\nu_\mathrm{s}}dz
\end{equation}
where $H(z)$ is the Hubble parameter at redshift $z$, $dE_\mathrm{s}/df_\mathrm{s}$ is the energy spectrum from a single inspiraling source evaluated in the source frame. We use the waveform template IMRPhenomD \cite{Husa:2015iqa,Khan:2015jqa} to compute $dE_s/df_s$ and integrate Eq.\ref{eq:omegagw} up to $z=20$; the contributions from higher redshifts can be neglected.

We assume a delta mass distribution for PBH binaries and have verified its stochastic background spectrum follows a power law with slope index $2/3$ within the LIGO/Virgo's most sensitive frequency band ($\sim20-100$ Hz). 
Therefore, we apply the LVK O3 upper limit for a power law spectrum with index 2/3 that $\Omega_\mathrm{GW}(\nu=25\mathrm{Hz}) < 3.4\times10^{-9}$ for a log-uniform prior on $\Omega_\mathrm{GW}$ and $\Omega_\mathrm{GW}(\nu=25\mathrm{Hz}) < 1.2\times10^{-8}$ for a uniform prior \cite{KAGRA:2021kbb} to constrain the PBH mass fraction.
The result is shown in the left panel of \cref{fig:fpbh} accompanied by the results from our direct search.
As shown, the constraint is generally weaker than the direct search by one order of magnitude, but becomes comparable for $0.1\msun$ where the individual source search is less sensitive. The stochastic background from PBH mergers at high redshift complements the direct search which only probes the local Universe.

\section{Conclusions}

Subsolar mass black holes would establish the existence of a population of PBHs \cite{Sasaki:2018dmp,Carr:2020gox,Green:2020jor} or hint at new physics outside of the standard model of stellar evolution~\cite{Shandera:2018xkn,Kouvaris:2018wnh,Dasgupta:2020mqg}.  We have performed a broad search for compact-binary coalescences where at least one component is less massive than one solar mass; binary sources span from comparable mass to extreme mass ratios up to $10^4$. We use our null detection to set new upper limits on the rate of mergers and the implied fraction of dark matter composed of PBHs. The top candidates from our analysis along with the configuration files necessary to reproduce the search are available at \url{https://github.com/gwastro/subsolar-O3-search}.

We have investigated the constraints on PBH abundance from the non-detection of the direct search for individual sources and a search for a stochastic background from the superposition of unresolved sources~\cite{KAGRA:2021kbb}; these probe gravitational-wave sources from the local and high redshift Universe, respectively. The observation of binary black hole mergers at high redshift is a key way to distinguish primordial from stellar origin because the former can merge before star formation \cite{Koushiappas:2017kqm,Chen:2019irf,Ng:2021sqn}.

Assuming a non-detection, we expect the next observing run (O4), with expected data available at the end of 2024, will be able to improve constraints by another factor of $2-4\times$~\cite{Aasi:2013wya}.
The future direct detection of subsolar mass black hole binaries or an excess in the merger rate at high redshift would give decisive evidence for PBHs. 

\vspace{1cm}
\acknowledgments

 We acknowledge the Max Planck Gesellschaft and the Atlas cluster computing team at AEI Hannover for support.  This research has made use of data, software and/or web tools obtained from the Gravitational Wave Open Science Center (https://www.gw-openscience.org), a service of LIGO Laboratory, the LIGO Scientific Collaboration and the Virgo Collaboration. LIGO is funded by the U.S. National Science Foundation. Virgo is funded by the French Centre National de Recherche Scientifique (CNRS), the Italian Istituto Nazionale della Fisica Nucleare (INFN) and the Dutch Nikhef, with contributions by Polish and Hungarian institutes.

\bibliography{references}

\end{document}